Plasma-photonic spatiotemporal synchronization of relativistic electron and laser beams

P. Scherkl[1,2*], A. Knetsch[3], T. Heinemann[1,2,3,4], A. Sutherland[1,2,5], A. F. Habib[1,2], O. S. Karger[4,] D. Ullmann[1,2], A. Beaton[1,2], G. Kirwan[1,2], G. G. Manahan[1,2], Y. Xi[6], A. Deng[6], M. D. Litos[7], B. D. O'Shea[5], S. Z. Green[5], C. I. Clarke[5], G. Andonian[6,8], R. Assmann[3], D. A. Jaroszynski[1,2], D. L. Bruhwiler[9], J. Smith[10], J.R. Cary[7,11], M. J. Hogan[5], V. Yakimenko[5], J. B. Rosenzweig[6], and B. Hidding[1,2]

*paul.scherkl@strath.ac.uk

**Affiliations**

[1] Scottish Universities Physics Alliance, Department of Physics, University of Strathclyde, Glasgow, UK.

[2] The Cockcroft Institute, Daresbury, UK.

[3] Deutsches Elektronen-Synchrotron DESY, Hamburg, Germany.

[4] Department of Experimental Physics, University of Hamburg, Hamburg, Germany.

[5] SLAC National Accelerator Laboratory, Menlo Park, California, USA.

[6] Department of Physics and Astronomy, University of California Los Angeles, USA.

[7] Center for Integrated Plasma Studies, Department of Physics, University of Colorado, Boulder, Colorado, USA.

[8] Radiabeam Technologies, Santa Monica, CA 90404, USA.

[9] RadiaSoft LLC, Boulder, CO 80301, USA.

[10]Tech-X UK Ltd., Daresbury, UK.

[11]Tech-X Corporation, Boulder, USA.

**Abstract**

Modern particle accelerators and their applications increasingly rely on precisely coordinated interactions of intense charged particle and laser beams. Femtosecond-scale synchronization alongside micrometre-scale spatial precision are essential *e.g.* for pump-probe experiments, seeding and diagnostics of advanced light sources and for plasma-based accelerators. State-of-the-art temporal or spatial diagnostics typically operate with low-intensity beams to avoid material damage at high intensity. As such, we present a plasma-based approach, which allows measurement of both temporal and spatial overlap of high-intensity beams directly at their interaction point. It exploits amplification of plasma afterglow arising from the passage of an electron beam through a laser-generated plasma filament. The corresponding photon yield carries the spatiotemporal signature of the femtosecond-scale dynamics, yet can be observed as a visible light signal on microsecond-millimetre scales.

**Introduction**

The experimental use of intense, femtosecond-duration particle beams together with similarly short laser pulses becomes increasingly important in exciting and probing ultrafast processes[1,2,3] occurring in plasma, molecular, atomic, and nuclear structures. For example, modern light sources such as free-electron-lasers require precise spatiotemporal control of laser-electron beam interactions for seeding[4] and diagnosis[5]. Inverse Compton scattering acting as a light source[6,7,8,9] as well as a beam diagnostic[10,11] has similar demands. An emerging class of high field plasma accelerators[12,13,14,15,16] also strongly benefits from the spatiotemporal coordination of electron and laser beams for driving plasma waves, control of injection processes, and (staged) acceleration of high-quality particle beams. This demand, extending across multiple fields of research, has given rise to a broad range of synchronization and alignment techniques, such as electro-optic sampling[17,18] (EOS), beam time-of-arrival monitoring measurement techniques based on excited cavities[19], coherent transition radiation[20], cross-correlated THz radiation from undulators[21], and various schemes in free electron lasers[22]. These techniques individually offer either measurement of temporal synchronization or spatial alignment of electron and laser beams. To measure both, combinations of diagnostics must be implemented separately along the beamline, *e.g.* by using EOS for temporal overlap measurement at one position and optical transition radiation beam profile monitors for spatial overlap at another one. Further, damage thresholds prohibit the use of intercepting diagnostics for most intense and focused laser and electron beams, such that they must be employed away from the beam focus and/or at reduced beam power levels. The same reasons prevent application of these diagnostics in the presence of plasma.

Addressing the demand for measuring spatial and temporal coordination directly at the interaction point and the current limitations outlined above, we report on a novel plasma-based approach allowing for precise synchronization and spatial alignment of intense laser and electron beams. This method provides an online diagnostic applicable in conventional and plasma-based accelerator schemes.

**Results**

**Experimental setup** Figure 1a shows the layout of experiments conducted at the Facility for Advanced Accelerator Experimental Tests (FACET) at the SLAC National Accelerator Laboratory[23], where a focused ultra-relativistic 20 GeV electron beam with root-mean-square (r.m.s.) length of $\sigma_z \approx 64$ µm propagates through a mixed $H_2$ and He gas reservoir at sub-atmospheric pressure of 5.3 mBar. The setup employs a Ti:Sapphire laser pulse with duration of $\tau_L \approx 60$ fs full-width-at-half-maximum (FWHM). Several metres upstream of the electron beam focus a collimated, low intensity laser fraction split-off from the source laser pulse is superimposed with the electric field from the electron beam on an EOS crystal. It provides relative time-of-arrival measurement and, being an established diagnostic, benchmarks the new effects we describe here. The main laser pulse arm is focused perpendicularly to the path of the electron beam to a spot size of $w_0 \approx 38$ µm (FWHM) in the immediate vicinity of the electron beam focus position. By tuning the energy of this laser pulse to yield intensities in the range $I \approx 10^{14} - 10^{15}$ Wcm$^{-2}$ in the gas, a millimetre-long $H_2$/He cold plasma filament of width $d_{plasma} \approx 100$ µm is generated by tunnelling ionization[24]. A CCD camera monitors the interaction region and integrates radiation from the characteristic He plasma afterglow line[25] $\lambda_{He} \approx 587$ nm. The purely laser-generated plasma afterglow signal results from hydrodynamic expansion and subsequent recombination and relaxation processes[26] in absence of the electron beam and is shown in Fig. 1b. The electron beam alone, in contrast, does not generate plasma: at a peak electron density of $n_{b0} = N_b / (2\pi)^{3/2} \sigma^2_x \sigma_z \approx 3.5 - 5 \times 10^{16}$ cm$^{-3}$ its corresponding radial electric fields $E_r(r) = [N_b e / (2\pi)^{3/2} \varepsilon_0 \sigma_z r] [1 - \exp(-r / 2\sigma^2_x)]$ are in the multi-GV/m range, below the tunnelling ionization threshold for hydrogen[27]. Here, $r$ is the radial coordinate, $N_b$ is the number of beam electrons, $\sigma_x$ is its transverse r.m.s. beam size, $e$ is the elementary charge and $\varepsilon_0$ the vacuum permittivity. Impact ionization of the gas by the transient electron beam is negligible due to the low cross sections associated with highly relativistic energies[28,29]. As such, if the electron beam is switched off, is substantially misaligned with respect to the laser pulse axis, or passes before the laser pulse, only the laser-generated plasma afterglow signal is observed.

If, however, the electron beam intersects the laser path after the laser pulse has passed, it interacts with the thin laser-generated plasma filament of electron density $n_e \approx 1.9 \times 10^{17}$ cm$^{-3}$ ($n_{b0} \approx 0.1\, n_e$) instead of neutral gas. Then, the unipolar transverse electric field of the focused electron beam briefly couples with this seed plasma filament for the duration of their geometrical overlap, and transfers energy into the plasma. This initial kick accelerates plasma electrons to moderate, non-relativistic, broad-band velocities with peak energies $W_{max} \approx \pi [eE_{peak} \sigma_z / \beta c]^2 / m_e \approx$ 100's of keV and expels them partially from the initial plasma volume. Here, $E_{peak}$ denotes the peak radial electric field of the beam, $m_e$ is the electron mass, and $\beta c$ the velocity. The interaction of electron beam and plasma filament distinctively changes the observed afterglow signal, as shown in Fig. 1c, by increasing its volume and intensity substantially compared to the signal displayed in Fig. 1b.

**Particle-in-cell simulations** Figure 2 shows the corresponding plasma dynamics modelled in three-dimensional (3D) particle-in-cell (PIC) simulations using VSim[30] (see Methods). The strong electric fields of the electron beam are rapidly screened inside the plasma, so that significant energy transfer occurs only if beam and plasma volumes overlap. In this case, direct heating and expulsion of laser-ionized plasma electrons in the immediate range of the electron beam is accompanied by electron plasma density waves which propagate along the filament axis. They spread the beam-induced perturbation outwards, beyond the direct beam field-excited region, up to the limits of the millimetre-long plasma column. The associated electromagnetic surface waves are of the same order of magnitude as the electron beam electric fields in vacuum, and further heat the plasma electrons which are not directly affected by the beam fields (see Supplementary Video 1). Because of the attracting potential set up by the plasma ions, plasma electrons expelled from the initial volume perform complex oscillations around the filament (see selected trajectories depicted in Fig. 2) which evolve due to the combined effects of initial kick, plasma density waves, and electromagnetic fields sustained by the plasma. These oscillations are inherently anharmonic, because the nominal plasma wavelength $\lambda_p = 2\pi c/[(n_{He}+2n_{H2})e^2/m_e\varepsilon_0]^{1/2} \approx 77$ µm is of the order of the initial plasma diameter. In this scenario, a large fraction of plasma electrons periodically propagates through the gas surrounding the original plasma. These electrons subsequently ionise the ambient gas via impact ionisation as they periodically exhibit instantaneous kinetic energies $W_{kin} > W_{thresh, H2} \approx 15.4$ eV and $W_{kin} > W_{thresh, He} \approx 24.6$ eV[25] corresponding to very large impact ionization cross sections $\sigma_{cs}$. Over time, the plasma electrons therefore create and accumulate a substantial additional plasma volume due to their orbits passing through the surrounding neutral gas. The spectral evolution of the plasma electrons inside and outside of the fully ionized plasma region is shown in Fig. 3 and Supplementary Video 2. While the spectral distribution remains nearly constant over extended timescales, the kinetic energies of individual electrons change rapidly during their oscillations (selected electrons are highlighted in Fig. 3 to illustrate this microscopic feature). On much longer time scales, the complex plasma dynamics excited by the electron beam further cause ion motion and expand the initial volume substantially compared to the hydrodynamic case. The result of this expansion can be seen by comparing Fig. 1 B and C, as the afterglow intensity is distributed well beyond the initial laser-only boundaries.

Summarising, the kinetic energy originating from the fast initial energy transfer from the electron beam to the seed plasma filament is thus gradually transferred into the creation of additional plasma. Eventually, recombination and atomic relaxation processes produce the emissions of the plasma afterglow[31,32,33,34], which exhibits intense lines in the optical spectrum for helium and can be observed by the CCD camera.

**Experimental measurements** We apply these findings to measure spatial alignment and time-of-arrival between electron beam and laser pulse. In both the spatial and temporal measurement modes, changing the geometric overlap varies the fraction of the beam fields transferring energy into the plasma, and thus the intensity of the resulting afterglow amplification.

The first measurement mode scans spatial alignment with fixed relative time-of-arrival (TOA) $\Delta t_{delay} = TOA_{laser} - TOA_{e\text{-beam}} = 2.1$ ps between laser and electron beam to ensure interaction of the electron beam with a fully formed laser-generated seed plasma. By introducing translations $\Delta y$ between electron beam and laser axis, the radial electric field of the electron beam intersects to a lesser degree with the pre-formed plasma region, thus reducing the total amount of energy transferred. Consequently, the integrated afterglow signal peaks for central overlap $\Delta y = 0$ and decreases with larger offsets $|\Delta y|>0$. Eventually, the amplification ceases completely, yielding only the afterglow from the unperturbed filament. Fig. 4 presents a $\Delta y$-alignment scan that reveals a Gaussian distribution of amplified plasma afterglow with characteristic width $\sigma_y = 64.8$ µm r.m.s. This curve agrees well with PIC simulation results and exhibits a linear relation between the energy transferred into the plasma and the afterglow amplification in the given interaction regime. Comparing experiment and simulation, the central peak position corresponds to optimal overlap of electron beam and plasma axes and can, therefore, readily be established within the accuracy of the imaging system. From the 99 shots in the given dataset, the central overlap between beam and plasma can be determined within 4.1 µm accuracy, which is substantially smaller than the dimensions of electron beam and plasma. This plasma afterglow response constitutes a multi-shot alignment diagnostic with accuracy dependent on the number of shots per setting. If, furthermore, the dependency shown in Fig. 4 is sufficiently well characterized either from experiment or theory, it represents a gauge curve for the given electron and laser beams. In this case, the method can quantify the absolute alignment between electron and laser beam on a single-shot basis. To improve the accuracy, it is desirable that the sources of parametric jitter, *e.g.* laser and electron beam size and duration, energy and charge, intensity and beam density, respectively, are well known, can be co-recorded and/or are substantially reduced.

In the second measurement mode, the spatial translation is fixed and the timing between electron and laser beam is varied. Setting the spatial overlap to the central position $\Delta y = 0$, a delay scan in a series of consecutive shots from -2.5 ps $< \Delta t_{delay} <$ 4.0 ps reveals a strong dependence between TOA and afterglow signal, as shown in Fig. 5. The signal drops sharply when the electron beam arrives at the same time or earlier than the laser pulse because the seed plasma volume is not yet fully formed or not present at all. As result, the scan yields a sigmoid transition with a width of ~315 fs (r.m.s.) between zero and maximal afterglow amplification. This region allows to determine the degree of temporal coincidence between laser and electron beam. On the steep quasi-linear transition region around $\Delta t_{delay} \sim 0$, the afterglow signal reacts sensitively to small changes in the TOA and therefore allows for determining the synchronisation with high resolution. In the given experimental setup, the raw data yields an all-optical TOA synchronization measurement from the quasi-linear transition region with shot-to-shot accuracy of 55 fs. This measurement represents a multi-shot TOA diagnostic, and its shot-to-shot accuracy can be further improved by increasing the number of shots per setting as discussed earlier. As in the alignment measurement, the experimentally achieved TOA accuracy also results from the shot-to-shot variations of electron beam and laser pulse parameters and relative alignment jitter between laser and electron beam. Most critically, the system-inherent shot-to-shot TOA jitter between electron beam and laser beam of ~109 ± 12 fs r.m.s. has large influence on the obtainable accuracy. This jitter results to a significant degree from the strong longitudinal compression of the electron beam in the magnetic chicanes of the SLAC linac. Calibrating the data with EOS time stamps partially removes the effect of this shot-to-shot TOA jitter, and enhances the combined TOA resolution to 16 fs. It shall be noted that while this first experimental demonstration benefits strongly from an EOS benchmarking the relative TOA jitter, the amplified plasma afterglow can serve as standalone diagnostic and uniquely offers absolute TOA measurements at the interaction point. We find good agreement between measurement and simulation, and the sigmoid shape of the temporal dependence curve supports the hypothesis of a linear relation between transferred energy and afterglow amplification in the parameter regime explored (see Supplemental Figure 1). This relationship implies that the plasma effectively samples the beam envelope, which allows for precise TOA determination. As discussed above, characterizing the gauge curve shown in Fig. 5 sufficiently precise can allow for single-shot TOA measurements provided that other sources of parameter jitter are low or measured independently.

The sensitivity and multi-parameter dependency of the amplified plasma afterglow thus enables a versatile diagnostic which can measure various parameters of the interaction. For example, in laser-early-mode, the setup can operate as diagnostic for electron beam duration (Suppl. Fig. 1, 2) and radius (Suppl. Fig. 3), respectively.

**Discussion**

The transient overlap of electron beam fields with a laser-generated plasma filament can be exploited as highly sensitive and versatile diagnostic system. The laser-gated seed plasma acts as spatiotemporally amplifying detector medium. The initial femtosecond-micrometre-scale interaction is transformed through localized plasma dynamics, eventually manifesting as amplified plasma afterglow light emission – an observable on the microsecond and millimetre scale in the visible spectrum. We have used this approach for synchronization and alignment of intense electron and laser beams directly at their interaction point with a single setup. Here, their intensities exceed the permissible measuring range and damage thresholds of many conventional diagnostics, while plasma as already ionized medium is sensitive to varying interaction parameters at such intensity levels, while not being destroyed in the process. In the context of these measurements, this capability was exploited in a first application to facilitate a plasma photocathode injection experiment[35], which relies on precise spatiotemporal alignment and synchronization on the micrometre and femtosecond-scale to release electrons inside a small fraction of a ~100 µm long plasma wave via tunnelling ionization.

The plasma afterglow-based measuring process is minimally intrusive and highly efficient. Both the energy required for generation of the seed plasma as well as the energy transferred by the electron beam to the seed plasma are of the order of 1 mJ, corresponding to ~0.001% of the total electron beam energy at FACET. This suffices to boost the afterglow signal by approximately a factor of ~50 during a spatiotemporal overlap scan.

Benchmarking of the data to predictions of PIC simulations shows that the experimentally observed afterglow amplification scales linearly with the total energy transferred by the beam into the plasma. In this study, we operate in the linear (overdense) beam-plasma interaction regime, as $n_{b,0}/n_e < 1$. The electron beam duration is longer than the width of the plasma filament; both attributes of the interaction ensure a quasi-adiabatic plasma response. At underdense working points, the energy transfer into plasma may significantly change the plasma dynamics and subsequent recombination rates depending on the degree of nonlinearity of the interaction[36]. One must still verify that the linear relation between initially transferred energy and plasma afterglow amplification holds in these regimes.

The obtainable resolution of the technique depends on the dynamic range of the experimental detector setup, the maximum afterglow amplification, and the gradient of the transition. The latter, in turn, profits from rapid Debye screening of the electric field of the electron beam inside the plasma due to its transverse size $\sigma_{x,y} \gg \lambda_D$, with $\lambda_D = (\varepsilon_0 k_B T/n_e^2)^{1/2}$ being the Debye length, $k_B$ denotes Boltzmann's constant and $T$ the electron temperature. Therefore, the coupling volume inside the filament is confined, and allows the plasma volume to sample the beam precisely. Simulations (see Suppl. Fig. 1) show that the transitions can become substantially steeper for shorter and narrower electron beams such as those expected *e.g.* from FACET-II, or from state-of-the-art plasma-based accelerators. For FACET-II, for example, the transition slope for a comparable parameter regime is expected to steepen by a factor of 4 compared to current measurements at FACET, thus improving resolution equivalently. Another, simple approach improving the resolution on the transition region decouples the amplifying neutral gas from the seed plasma. For example, reducing the laser energy can avoid ionization of helium such that only the laser-generated hydrogen filament receives energy from the electron beam. In this case, helium only gets ionised by heated hydrogen electrons and secondaries. As this procedure allows for much higher helium densities, the signal-to-noise-ratio can be substantially increased.

Development of a closed analytic description of electron-beams partially overlapping bounded plasmas, the complex long-term dynamics and their influence on the afterglow signal is beyond the scope of this work and will be subject to further research. Particularly, investigating the plasma afterglow amplification spatially and temporally resolved could identify and quantify the individual processes of energy conversion between the electronic dynamics investigated by PIC and the recombination timescale of the heated plasma.

As shown by measurement and simulations, the amplified afterglow signal is notably sensitive to various interaction parameters: electron beam duration, size, and charge; laser beam intensity and power; and gas density. This multi-parameter sensitivity promises further optimized resolution and represents a valuable attribute of the approach as it enables broad versatility. On the other hand, the latter in turn means that the resolution of the quantity to be measured is susceptible to the shot-to-shot jitter of other interaction parameters. Full exploitation of the versatility and sensitivity of this method therefore requires, as is true of many diagnostics, minimization of jitter contributions, and/or their simultaneous measurement with auxiliary diagnostics. The former is a general goal of any particle accelerator and is subject to intensive development, *e.g.* in context of free-electron-laser facilities. The latter can be addressed by simultaneous composite measurements. While in our proof-of-concept experiments we have exploited the plasma afterglow in two different measurement modes for two different series of shots sequentially, these modes can be applied to the same electron beam shot by generating multiple seed plasma filaments side by side considering the negligible energy loss in each filament. Such composite measurements could isolate and deconvolve the parametric dependencies, yielding a comprehensive characterization of laser beam, electron beam, as well as their spatiotemporal relation. At advanced stability and reduced shot-to-shot jitters as attainable at modern photocathode-based facilities such as free-electron-lasers, sub-femtosecond and sub-micrometre resolution may be ultimately obtainable.

We emphasize the applicability of the effect for intense and focused beams, as plasma-based techniques are not limited by conventional damage thresholds, and its applicability directly at the interaction point of such beams and even in the presence of plasma. This opens numerous experimental doors *e.g.* for pump-probe experiments, for the optimization of seeded free-electron-lasers, inverse Compton scattering light sources, advanced and staged plasma accelerator schemes, and beam-laser experiments probing quantum electrodynamics.

**Materials and Methods**

**Experiment** The laser-plasma-electron beam interaction experiment at FACET is conducted within a large gas volume separated from the SLAC beamline vacuum. The homogeneously distributed gas mixture consists of $H_2$ and He in 1:1 ratio at pressure $p =$ 5.3 mBar corresponding to hydrogen and helium gas densities of $n_{H2} \approx n_{He} \approx 6.3\times10^{16}$ cm$^{-3}$. The SLAC linac provides an ultra-relativistic electron beam at 1 Hz repetition rate with energy $W \approx 20.4$ GeV $\pm$ 2 % (FWHM), total charge $Q = 3.0$ nC $\pm$ 0.6% (r.m.s), length $\sigma_z \approx$ 64 µm $\pm$ 1.9 % (r.m.s.) and transverse size $\sigma_x \approx 22$ µm and $\sigma_y \approx 44$ µm (r.m.s.). The laser pulse is compressed to a FWHM duration of $\tau_L \approx 60$ fs and its energy is adjusted to 4.9 $\pm$ 0.1 mJ by polarized beam splitters. An *f*/22.9 off-axis parabola (OAP) mounted on a motorized stage with five degrees of freedom focuses the laser to $w_0 \approx 38$ µm (FWHM) and provides a mean intensity $I \approx 3\times10^{15}$ Wcm$^{-2}$ to generate a plasma filament perpendicular to the electron beam axis (see Fig. 1 b). A CCD installed along the laser propagation direction measures the spatial alignment between electron beam and laser pulse. To achieve precise overlap of laser and electron beam propagation axes, a motorized optical transition radiation (OTR) screen placed at the interaction point yields the electron beam position and allows alignment of the laser centroid with the OTR screen removed. The electron orbit in *y*-direction varies with an average shot-to-shot jitter of 6.6 µm (r.m.s.) characterized by ~100 shots observed on an OTR screen and the laser pointing in this direction jitters by 7.6 µm (r.m.s). The relative time-of-arrival between electron beam and laser pulse is tuned by a linear stage with ~333 ps delay range. A second CCD (see Fig. 1 b) images the recombination/atomic relaxation radiation originating from the interaction with 25 ms integration time, using a spectral band pass filter of range 589.3 $\pm$ 5.0 nm. A low-intensity pulse split off before the OAP is used to measure the relative system-inherent electron beam-laser pulse synchronization jitter for every shot by electro-optic sampling. The electro-optic sampling (EOS) consists of a 100 µm thick gallium phosphide crystal, placed a few millimetres from the electron beam orbit and irradiated by the laser at an angle of 45°. The signal is analysed by a polarization filter and imaged onto a CCD camera. The resulting image is background-subtracted and calibrated by altering the path length of the laser-beamline averaged for every step over the intrinsic timing jitter. This EOS maps the relative TOA between laser and electron beam to physical delays. The EOS temporal calibration is $\tau_{res} = 25.8 \pm 2.5$ fs px$^{-1}$. Furthermore, the EOS measures the relative TOA jitter between electron beam and laser pulse as 109 $\pm$ 12 fs r.m.s [37].

**Data analysis** The experimental timing data shown in Fig. 5 correlates each image produced by the *z-x*-CCD with an individual relative TOA determined by the EOS unit, which accounts for temporal jitter between laser and electron beam. Projections of recorded plasma afterglow images are then sorted by EOS TOA. Integrating all pixels per shot and subsequent normalization yields the red scatter plot for the delay scan in Fig. 4. We model the corresponding transition with a sigmoid function $s(t) = a / (1 + \exp((t - t_0) \cdot k))$. The average camera background noise level of 707 counts limits the resolution of our imaging system. Error propagation of the detector function $s(t)^{-1}$ yields the TOA accuracy of the experiment for raw data (54.7 fs) and for the data set calibrated by EOS time stamps (15.8 fs).

For the alignment scan, the collected signals are correlated with the vertical focus position determined by the centroid of the imaged focus. The integrated counts per shot are fit by a bi-Gaussian distribution $f(y) = a_1 \cdot \exp(-((y-b_1)/c_1)^2) + a_2 \cdot \exp(-((y-b_2)/c_2)^2)$. The first term describes the amplified plasma afterglow signal from the interaction with the electron beam and agrees well with PIC simulations. The second one results from systematic aberration errors due to movements of the OAP which reduces the plasma volume with increasing misalignment. In Fig. 4 we subtract the aberration from the signal to make simulation and experiment comparable. From the amplified signals detector function $f(y)^{-1}$ we obtain best measured accuracy of 4.1 µm at the peak of the curve from error propagation.

**3D particle-in-cell simulations** The experimental conditions in the central area of the beam-plasma interaction are modelled in a fully explicit 3D particle-in-cell simulations using VSim[30] with a simulation box size 1.9 mm × 1.6 × 0.8 mm (1.2 mm × 1.0 mm × 1.4 mm for the alignment scan in Fig. 5) and 3 µm cell size in each direction. This corresponds to 633 × 533 × 266 cells (399 × 333 × 466 for the alignment scan) filled with 8 nominal macroparticles per cell for the plasma and 16 nominal macroparticles per cell for the electron beam. This grid resolution is designed to capture the macroscopic plasma response to the beam field as well as to resolve the electromagnetic dynamics of accelerated plasma electrons with reasonable computational effort. The electron beam is modelled as a bi-Gaussian distribution with the FACET parameters summarized in the experimental methods. The plasma filament inside this gas is generated by an 800 nm laser pulse with energy 4.9 mJ, waist $w_0 = 38$ µm (FWHM) and FWHM pulse duration $\tau = 60$ fs. The laser is implemented in the envelope approximation and ionizes via an ADK[27] tunnel ionization model. To study the interaction for different delays $\Delta t_{delay}$ between electron beam and plasma filament the laser pulse is launched at different times with focus centred on the beam axis or with transverse offset at fixed timing (laser 1.0 ps ahead of electron beam emulates the fully developed plasma for finite box size) for the alignment scan. In Fig. 4 and 5, the total energy transferred from the electron beam into the plasma is evaluated for each delay and alignment. The maximal energy transferred into the plasma amounts to 0.93 mJ, a negligible fraction of the total beam energy of ~67 J. Since as EOS time stamps do not provide absolute timing information, the measured integrated counts in in Fig. 4 D are shifted on the TOA axis such that the curve agrees with the simulated data.

**References and Notes**

**Acknowledgments**

**Funding:**
This work was performed in part under US DOE Contracts DE-SC0009914 (UCLA)., DE-AC02-76SF00515 (SLAC) and DE-SC0009533 (RadiaBeam Technologies). BH acknowledges support by the DFG Emmy-Noether program. We acknowledge further support by H2020 EuPRAXIA (Grant No. 653782), UK EPSRC Grants No. EP/N028694/1, EC Laserlab-Europe (Grant No. 284464), EuCARD-2 (Grant No. 312453). This work used computational resources of the National Energy Research Scientific Computing Center, which is supported by DOE DE-AC02-05CH11231, of JURECA (Project hhh36), of HLRN, and of Shaheen-II (Project k1191). D.L.B. acknowledges DOE contract DE-SC0013855.

**Author contributions:**
P.S., A.K., T.H., A.S. and A.F.H. contributed equally to this work. A.K. and B.H. conceived the idea. P.S. and B.H. led the paper writing. The experiment was done by A.K., P.S., T.H., O.S.K., G.G.M., Y.X., and A.D. Simulations and analysis were performed by P.S., T.H., and A.S., supported by D.U., A.F.H, G.K. and A.B. The experiment was strongly supported by M.D.L., B.D.O., S.Z.G., C.I.C. Theoretical, numerical and computational aspects of this paper were contributed by G.A., R.A, D.A.J., D.L.B., J.S., J.R.C., M.J.H, V.Y. and J.B.R.

**Competing interests:**

The authors have no competing interests.

**Data and materials availability:**

Data associated with research published in this paper will be available at publication.

**Figures and Tables**

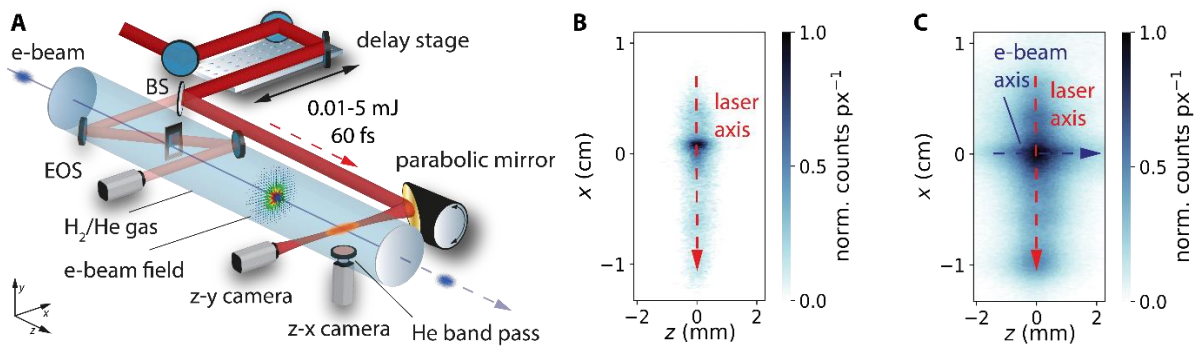

**Fig. 1. Experimental layout and key observables. (A),** Experimental setup at SLAC FACET. The electron beam propagates in *z*-direction through a $H_2$/He gas mixture and its transverse electric field signature is imprinted upon a laser pulse when it passes by a gallium phosphide (GaP) crystal for electro-optic sampling (EOS). A beam splitter (BS) sends the main laser arm downstream, where a parabolic mirror focuses it to produce a narrow plasma filament perpendicular to the path of the electron beam. The interaction point is monitored by two CCD cameras. **(B)**, Experimental measurement (*z-x* camera) of laser-generated plasma afterglow without interaction with the electron beam. A spectral bandpass filter isolates the He afterglow emission at ~587 nm. **(C)**, The plasma afterglow intensity and volume are substantially amplified if the electron beam interacts with the laser-generated plasma filament. Here, the electron beam arrives 2.2 ps after the laser pulse.

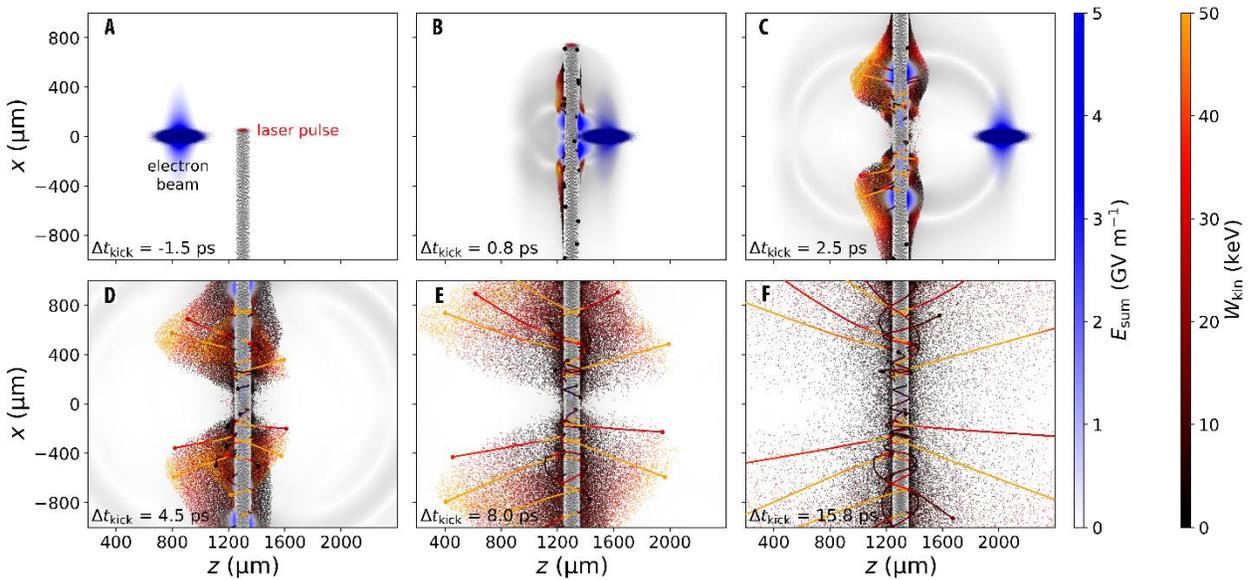

**Fig. 2. Snapshots of 3D particle-in-cell simulations. (A),** The electron beam (dark blue) propagates in $z$-direction while the laser pulse generates the perpendicular cold $H_2$/He plasma filament (black dots, projection in $z$-$x$ plane), and reaches the centre of the filament at $\Delta t_{kick} = 0$. **(B, C),** The electron beam crosses the filament and plasma electrons gain energies up to ~hundred keV by interacting with the unipolar beam field (blue). Subsequent plasma density waves with associated fields of several GV/m (blue, central slice of simulation) propagate along the filament and induce strong density perturbations. **(B-F),** Localized plasma electron oscillations visualized by typical electron trajectories (solid lines, projected to the $z$-$x$ plane) color-coded by momentary energy. These are anharmonic due to the finite plasma width. Oscillating electrons repeatedly leave the initial plasma volume and propagate through ambient neutral gas (not shown) on their orbits at eV to keV energies.

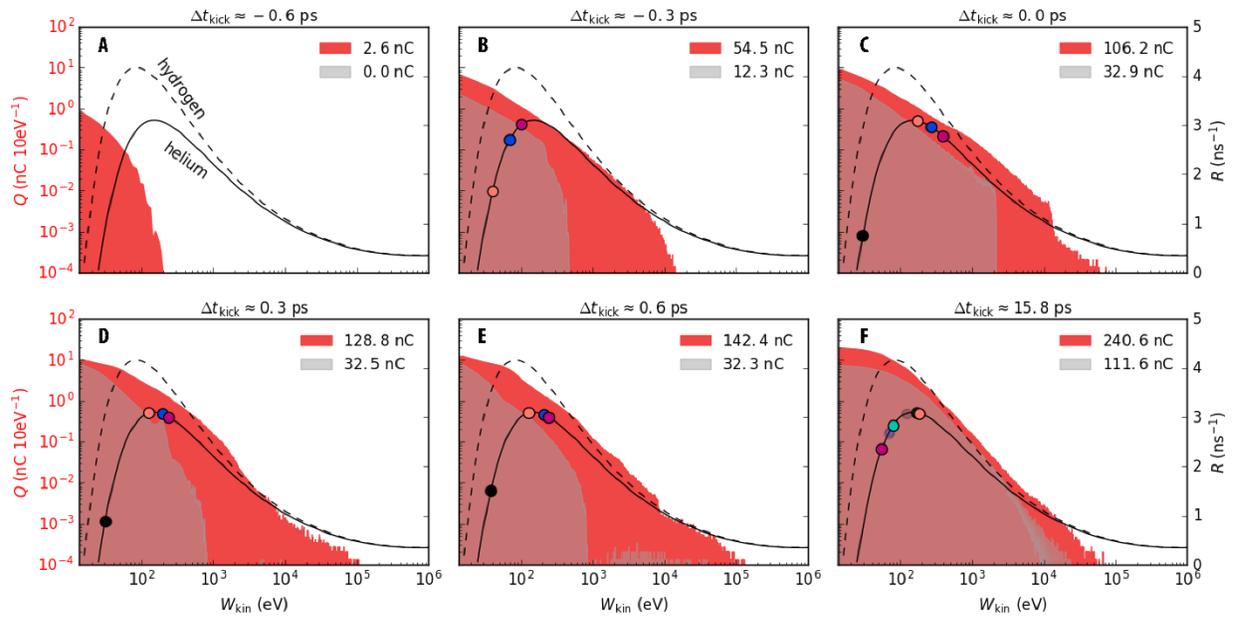

**Fig. 3. Spectral evolution of plasma electrons.** The histograms show the spectral distribution $Q$ of heated plasma electrons ($E_{kin} > W_{thresh, H2} \approx 15.4$ eV) presented in Fig. 2 outside (red) and inside (gray) the fully ionized plasma region with radius 24 µm. **(A-E),** During the initial interaction phase the plasma charge and electron energies quickly ramp up to a broadband, non-relativistic distribution as result of localized electron oscillations. **(A-F),** Coloured circles represent individual oscillating electrons shown in Fig. 2 to indicate their changing energies and corresponding momentary cross sections $\sigma_{Gas}$. Electrons inside (transparent colours) the initial plasma volume cannot collide with neutrals but do contribute to the re-distribution of energy along the plasma filament. Electrons outside (solid colours) ionize ambient neutral gas at significant impact ionization rates $R = n_{Gas}\, \sigma_{Gas}\, \beta c$ per electron (black lines) peaked at ~100 eV. Over time, those electrons transfer energy into formation of additional plasma and thus intensify the resulting plasma afterglow[32,34].

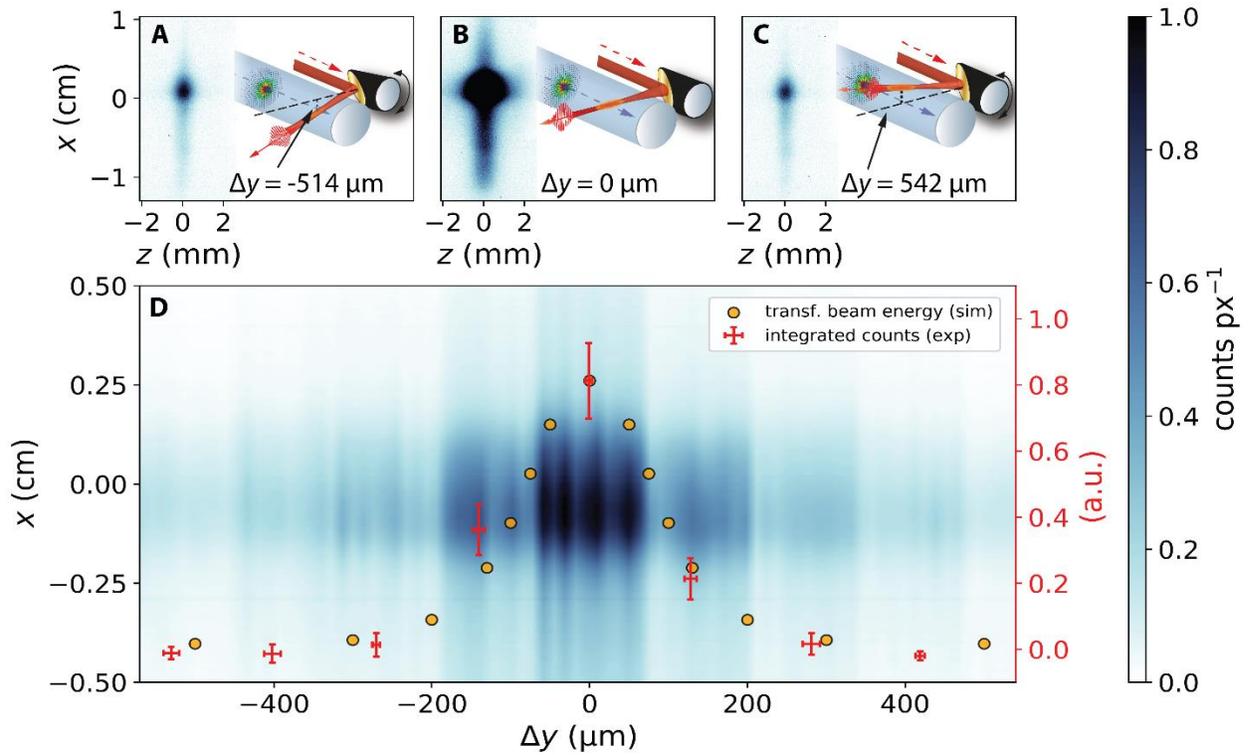

**Fig. 4. Experimental spatial alignment scan and PIC simulation**. **(A,C),** Shots with spatial misalignment between laser and electron beam axis reduce the maximum observed photon count measured for optimal alignment **(B)**. A fixed TOA $\Delta t_{\text{delay}} \approx$ -2.1 ps ensures a fully evolved plasma before electron beam arrival. **(D),** Complete alignment scan with 8-10 shots per setting. The raw images and integrated photon counts display a spatial Gaussian form with r.m.s. width $\sigma_y$ = 64.8 µm. The best overlap of beam and laser corresponds to the peak afterglow signal and can be determined with accuracy of 4.1 µm as result of the low number of shots. The simulated energy transferred to the plasma agrees well with the experimental curve.

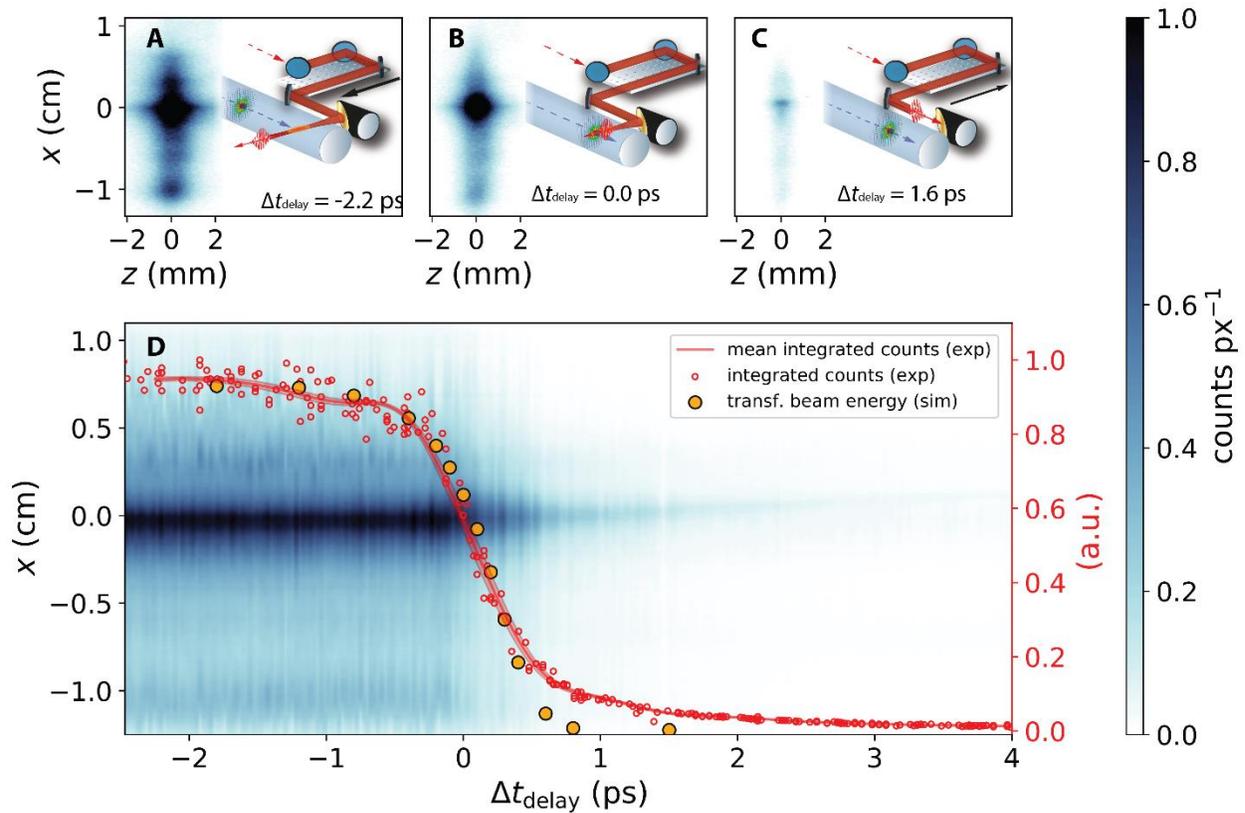

**Fig. 5. Experimental time-of-arrival scan and PIC simulation**. **(A),** The electron beam arrives after the laser pulse and interacts with the fully formed plasma filament producing maximum afterglow signal. **(B),** Both beams coincide, and the partially evolved plasma filament emits reduced signal. **(C),** The laser arrives after the electron beam, reducing the signal to the laser-only level. **(D),** The background raw data shows the complete transition over 256 consecutive shots within -2.5 ps $< \Delta t_{delay} <$ 4.0 ps, here calibrated by EOS time stamps. The normalized integrated photon count of those images exhibits a sharp transition allowing shot-to-shot TOA measurements with 16 fs accuracy (54 fs without EOS calibration). The data agrees well with the simulated energy transfer from electron beam into plasma, illustrating a linear dependence.